\documentclass[pra, aps, twocolumn, floatfix, showpacs]{revtex4-1}
\usepackage{graphicx, psfrag, amsmath, amssymb, times}

\begin{document}
\title{Topological superfluid phases of an atomic Fermi gas with in- and out-of-plane Zeeman fields and equal Rashba-Dresselhaus spin-orbit coupling}
\author{M. Iskin$^1$ and A. L. Suba{\c s}{\i}$^2$}
\affiliation{
$^1$Department of Physics, Ko\c c University, Rumelifeneri Yolu, 34450 Sar{\i}yer, Istanbul, Turkey. \\
$^2$Department of Physics, Faculty of Science and Letters, Istanbul Technical University, 34469 Maslak, Istanbul, Turkey. 
}
\date{\today}

\begin{abstract}
We analyze the effects of in- and out-of-plane Zeeman fields on the BCS-BEC evolution 
of a Fermi gas with equal Rashba-Dresselhaus (ERD) spin-orbit coupling (SOC).
We show that the ground state of the system involves novel gapless superfluid phases that 
can be distinguished with respect to the topology of the momentum-space regions 
with zero excitation energy. For the BCS-like uniform superfluid phases with 
zero center-of-mass momentum, the zeros may correspond 
to one or two doubly-degenerate spheres, two or four spheres, two or four concave 
spheroids, or one or two doubly-degenerate circles, depending on the combination of 
Zeeman fields and SOC. 
Such changes in the topology signal a quantum phase transition between distinct 
superfluid phases, and leave their signatures on some thermodynamic quantities.
We also analyze the possibility of Fulde-Ferrell-Larkin-Ovchinnikov (FFLO)-like 
nonuniform superfluid phases with finite center-of-mass momentum 
and obtain an even richer phase diagram.
\end{abstract}

\maketitle

\section{Introduction} 
\label{sec:intro}
While the topological matter and related phenomena have become a popular subject 
especially in the past few years within the condensed-matter community~\cite{hasan, sczhang, wen},
a new window has just recently been opened to the exotic world of topological
phases in atomic and molecular physics communities, with arguably a greater premise 
on a route toward studying them under highly controllable atomic 
settings~\cite{nistsoc, chinasocb, chinasocf, mitsoc}. 
The current advances in atomic systems offer the possibility 
of engineering both Abelian or non-Abelian artificial gauge fields on demand, 
by coupling the internal states of atoms to their center-of-mass motion via Raman 
dressing of atomic hyperfine states with laser fields. While there are a number 
of theoretical proposals for implementing atomic gases with various SOC symmetries, 
several experimental groups have so far only 
achieved an Abelian ERD coupling, first with Bose~\cite{nistsoc, chinasocb} and then with 
Fermi~\cite{chinasocf, mitsoc} gases. Allured by the experimental possibilities, 
there has been a growing interest in studying spin-orbit coupled atomic Bose 
and Fermi gases as functions of the tunable laser parameters including the strength 
and symmetry of the SOC field, $s$-wave scattering length and Zeeman fields. 
Both the uniform and trapped systems have been considered 
in all one, two and three dimensions, already revealing exotic single-, two-, few- 
and many-body properties at both zero and finite 
temperatures~\cite{zhai, shenoy, pu, galitski, gong, subasi, subasi2,wyi, carlos, zhang,liao,zhou,he,iskin}.

\begin{figure} [htb]
\centerline{\scalebox{0.38}{\includegraphics{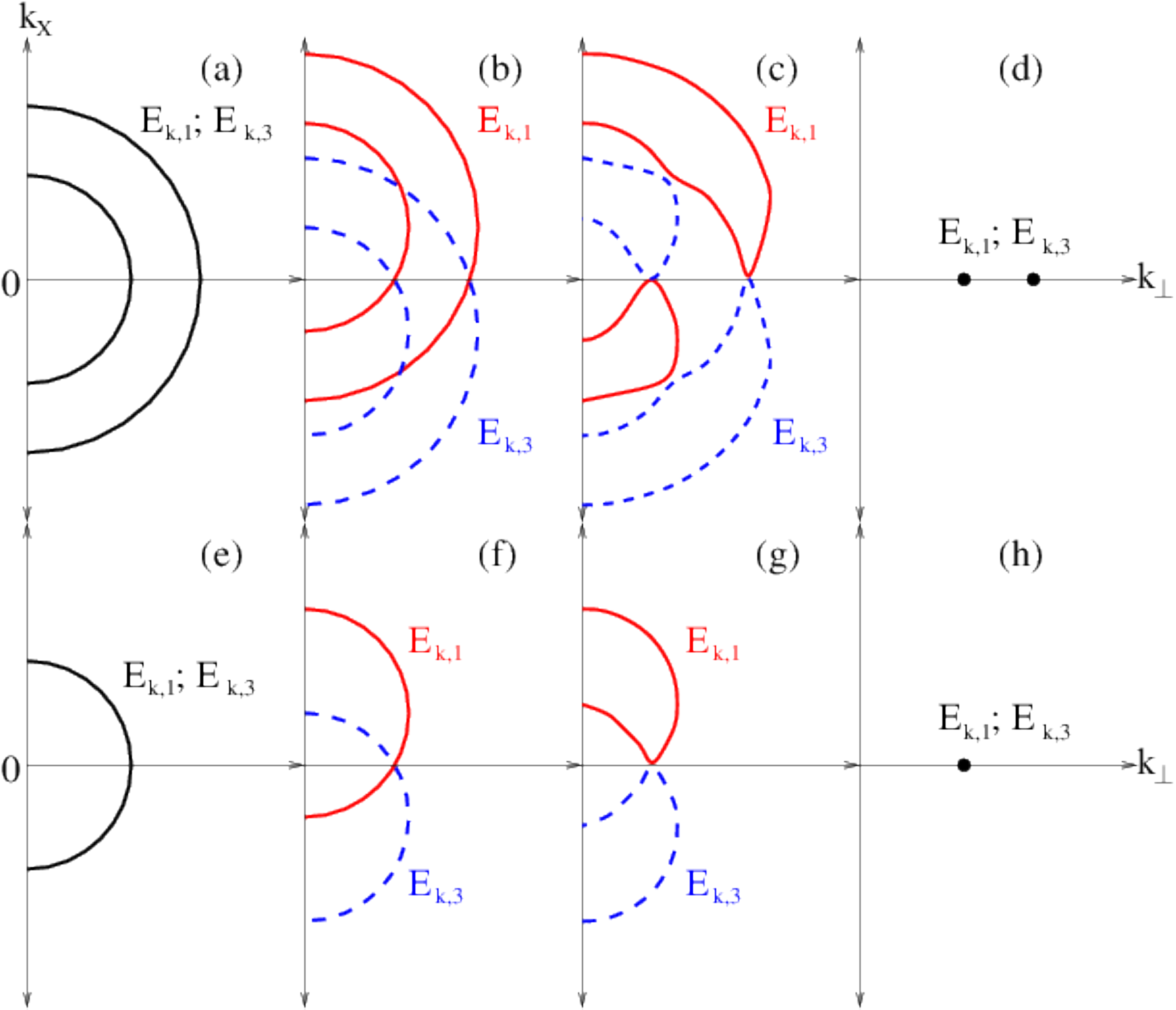}}}
\caption{
\label{fig:topclass} (color online)
The gapless superfluid phases can be classified with respect to the topology of 
their momentum-space regions with zero excitation energy. The top (bottom) figures 
illustrate the possible BCS-like uniform superfluid phases on the BCS (BEC) side, 
where the zero-energy regions correspond to
(a,e) two (one) doubly-degenerate spheres centered at $\mathbf{k} = \mathbf{0}$ 
when $\alpha = 0, h_O \ne 0$ and/or $h_I \ne 0$;
(b,f) four (two) spheres centered at $k_x = \pm M\alpha$ and $k_y = k_z = 0$ 
when $\alpha \ne 0, h_O = 0$ and $h_I \ne 0$;
(c,g) four (two) concave spheroids when $\alpha \ne 0, h_O \ne 0$ and $h_I \ne 0$; and
(d,h) two (one) doubly-degenerate circles centered at $k_x = 0$ when $\alpha \ne 0, h_O \ne 0$ and $h_I = 0$. 
Here, $\alpha$, $h_O$ and $h_I$ are the strengths of the SOC, out-of- and in-plane Zeeman fields, 
respectively. 
}
\end{figure}

In particular to the Fermi gases, depending on both the strength and the symmetry of the 
SOC field, it has been shown that the interplay between the SOC and out-of-plane Zeeman fields
gives rise to a variety of topologically-nontrivial and thermodynamically-stable 
gapless superfluid phases~\cite{gong, subasi, subasi2, wyi, carlos, zhang,liao,zhou,he,iskin}. 
In confined geometries, some of these phases have been 
shown to feature Majorana zero-energy bound states, for which the associated 
quasiparticle operators are self-Hermitian, i.e. a Majorana quasiparticle is its 
own anti-quasiparticle~\cite{iskin}. These quasiparticles play a key role in fault-tolerant 
quantum computation, and although they are predicted to be found in strongly-correlated 
systems in many fields of physics, including the fractional quantum Hall systems~\cite{moore}, 
chiral two-dimensional $p$-wave superconductors~\cite{volovik, read}/superfluids~\cite{tewari, mizushima}, 
three-dimensional topological insulator-superconductor heterostructures~\cite{fu},
one-dimensional nanowires~\cite{oreg, alicea}, 
spin-orbit coupled semiconductor-superconductor heterostructures~\cite{sau}, etc., 
their realization has proved to be quite difficult. 

Given that the cold atoms offer unprecedented control in comparison 
to the condensed-matter ones, there is a good chance of creating and observing 
Majorana quasiparticles with atomic systems in the near future. Motivated by 
these recent developments~\cite{chinasocf, mitsoc}, here we analyze the effects 
of not only the out-of-plane but also the in-plane Zeeman fields on the phase 
diagram of a Fermi gas with ERD coupling.
Since the inclusion of an in-plane Zeeman field complicates the many-body 
problem considerably even at the mean-field level, this term has been completely 
ignored in the recent literature~\cite{gong, subasi, subasi2, wyi, carlos, zhang, liao,zhou,he, iskin},
even though it appears naturally in the active experiments~\cite{chinasocf, mitsoc}. 
By including it here, we find gapless superfluid phases that can be distinguished 
with respect to the topology of the momentum-space regions with zero 
excitation energy. For instance, for the BCS-like uniform superfluid 
phases with zero center-of-mass momentum, 
the zeros may correspond to one or two doubly-degenerate spheres, 
two or four spheres, two or four concave spheroids, or one or two doubly-degenerate circles 
(see Fig.~\ref{fig:topclass}). These phases are distinct because, even though their 
order parameters all have the same $s$-wave symmetry, changes in the topology 
 leave strong signatures on some of the thermodynamic quantities, signaling 
 a topological quantum phase transition. We also analyze the possibility of 
 FFLO-like nonuniform superfluid phases with finite center-of-mass 
 momentum~\cite{FF,LO} and obtain an even richer phase diagram.

The rest of the paper is organized as follows. We introduce the Hamiltonian in 
Sec.~\ref{sec:mft}, and derive a complete set of nonlinearly-coupled 
self-consistency equations for the amplitude of the superfluid order parameter, 
center-of-mass momentum, total number and polarizations. 
We solve the resultant equations numerically and obtain the phase diagram 
of the system for the BCS-like uniform superfluid phases in Sec.~\ref{sec:bcs} 
and FFLO-like nonuniform superfluid phases in Sec.~\ref{sec:fflo}.
A brief summary of our main findings is given in Sec.~\ref{sec:conclusions}.

\section{Mean-Field Theory}
\label{sec:mft}
The results mentioned above are obtained by considering the following fields: 
the ERD coupling $-\alpha k_x \sigma_y$ which corresponds to a 
momentum-dependent Zeeman field in the in-plane $y$ direction, out-of-plane 
Zeeman field $-h_O \sigma_z$ and in-plane Zeeman field $-h_I \sigma_y$, 
where $\{ \alpha, h_O, h_I\} \ge 0$ are their strengths and $k_x$ is the 
$x$-component of the momentum. These fields have very different origin in
cold-atom systems, where the Raman coupling and detuning between the two 
laser beams driving the Raman transition from the two-photon resonance 
correspond to the out-of and in-plane Zeeman fields, respectively, and 
the ERD coupling is the Doppler shift an atom experiences 
as it moves in the two laser fields~\cite{nistsoc, chinasocb, chinasocf, mitsoc}.

The mean-field Hamiltonian for this system (in units of $\hbar = 1 = k_B$) can be written as,
$
H = H_0 + (1/2) \sum_{\mathbf{k}} \psi_\mathbf{k Q}^\dagger \textbf{D}_\mathbf{k Q} \psi_\mathbf{k Q},
$
where 
$
H_0 = \sum_{\mathbf{k}} \xi_\mathbf{-k_-} + |\Delta_\mathbf{Q}|^2/g
$
and the matrix $\mathbf{D}_\mathbf{k Q}$ is given by
\begin{align}
\label{eqn:ham}
\left( \begin{array}{cccc}
\xi_\mathbf{k_+} - h_O & S_\mathbf{k_+} + ih_I & 0 & \Delta_\mathbf{Q} \\
S_\mathbf{k_+}^* - ih_I & \xi_\mathbf{k_+} + h_O & -\Delta_\mathbf{Q} & 0  \\
0 & -\Delta_\mathbf{Q}^* & -\xi_\mathbf{-k_-} + h_O & -S_\mathbf{-k_-}^* + ih_I \\
\Delta_\mathbf{Q}^* & 0 &  -S_\mathbf{-k_-} - ih_I & -\xi_\mathbf{-k_-} - h_O
\end{array} \right).
\end{align}
Here,
$
\psi_{\mathbf{k Q}}^\dagger = 
[a_{\mathbf{k_+},\uparrow}^\dagger, a_{\mathbf{k_+},\downarrow}^\dagger,  a_{\mathbf{-k_-},\uparrow}, a_{\mathbf{-k_-},\downarrow}]
$
denotes the fermionic operators collectively with
$
\mathbf{k_\pm} = \mathbf{k \pm Q/2},
$
$\mathbf{Q} = (Q_x, Q_y, Q_z)$ is the center-of-mass momentum of the Cooper pairs,
$
\xi_\mathbf{k} = \epsilon_\mathbf{k} - \mu
$
is the shifted dispersion with $\epsilon_\mathbf{k} = k^2/(2M)$ and $\mu$ the chemical potential,  
$
\Delta_\mathbf{Q} = g\langle a_{\mathbf{k_+},\uparrow} a_{-\mathbf{k_-},\downarrow} \rangle
$
is the mean-field order parameter with $g$ the strength of the contact interaction and 
$\langle \cdots \rangle$ the thermal average, and $S_{\mathbf{k}} = i \alpha k_x$. 
We consider both BCS-like uniform superfluid phases with $\Delta_\mathbf{Q} = \Delta_0$, 
and FFLO-like nonuniform superfluid
phases with $\Delta_\mathbf{Q} = |\Delta_0| e^{i\mathbf{Q} \cdot \mathbf{R}}$, 
where $\mathbf{R}$ is the center-of-mass position. 
We eliminate $g$ in favor of the $s$-wave scattering length $a_s$ via the relation,
$
1/g = -M V/(4\pi a_s) + \sum_\mathbf{k} 1/(2\epsilon_\mathbf{k}),
$
where $V$ is the volume.
The thermodynamic potential for this Hamiltonian is given by,
$
\Omega = H_0 + (T/2) \sum_{\mathbf{k},\lambda} \ln \left[1-f(E_{\mathbf{k Q},\lambda}) \right],
$
where $T$ is the temperature, $\lambda = \lbrace 1,2,3,4 \rbrace$ labels the quasiparticle/quasihole 
excitation energies $E_{\mathbf{k Q},1} = -E_{-\mathbf{k Q},3}$ and $E_{\mathbf{k Q},2} = -E_{-\mathbf{k Q},4}$ 
as determined by the eigenvalues of the matrix given above, and
$
f(x)= 1/(e^{x/T}+1)
$ 
is the Fermi function.

The self-consistency equations for $|\Delta_0|$ and $Q_i$ are obtained by 
minimizing $\Omega$, i.e. $\partial \Omega / \partial |\Delta_0| = 0$ 
and $\partial \Omega / \partial Q_i = 0$ for $i \equiv (x,y,z)$, respectively, 
and this procedure leads to
\begin{align}
\label{eqn:gap}
\frac{|\Delta_0|}{g} &= -\frac{1}{4} \sum_{\mathbf{k},\lambda} \frac{\partial E_{\mathbf{k Q},\lambda}}{\partial |\Delta_0|} f(E_{\mathbf{k Q},\lambda}), \\
\label{eqn:Qi}
Q_i &= - \frac{2\pi^3 M}{k_0^3 V} \sum_{\mathbf{k},\lambda} \frac{\partial E_{\mathbf{k Q},\lambda}}{\partial Q_i} f(E_{\mathbf{k Q},\lambda}).
\end{align}
Here, while the momentum-space cutoff $k_0$ used in the numerical 
evaluation of $\mathbf{k}$-space integrations appears explicitly in Eq.~(\ref{eqn:Qi}), 
due to the term $\sum_\mathbf{k} 1 = V (k_0/\pi)^3$ in three dimensions where
$-k_0 \le k_i \le k_0$ for $i = \{x,y,z\}$, we checked 
that our results do not depend on its specific value as long as $k_0$ is sufficiently 
large compared to the Fermi momentum $k_F = (3\pi^2 N/V)^{1/3}$. 
In addition, we obtain the total number 
of particles and in- and out-of-plane number polarizations via 
$N = - \partial \Omega / \partial \mu$ and $P_{I, O} = -(1/N) \partial \Omega / \partial h_{I, O}$, 
leading to
\begin{align}
\label{eqn:ntot}
N &= \frac{1}{4} \sum_{\mathbf{k},\lambda} \left[ 1 - 2\frac{\partial E_{\mathbf{k Q},\lambda}}{\partial \mu} f(E_{\mathbf{k Q},\lambda}) \right], \\
\label{eqn:pI}
P_I &= -\frac{1}{2N} \sum_{\mathbf{k},\lambda} \frac{\partial E_{\mathbf{k Q},\lambda}} {\partial h_I} f(E_{\mathbf{k Q},\lambda}), \\
\label{eqn:pO}
P_O &= -\frac{1}{2N} \sum_{\mathbf{k},\lambda} \frac{\partial E_{\mathbf{k Q},\lambda}} {\partial h_O} f(E_{\mathbf{k Q},\lambda}).
\end{align}
We also use these expressions to extract the projections $n_{\mathbf{k}, \delta}$ 
of the momentum distribution along the out-of-plane $z$ direction where 
$\delta = (\uparrow, \downarrow)$, and in-plane $y$ and $x$ directions where 
$\delta = (\rightarrow, \leftarrow)$ and $\delta = (\swarrow, \nearrow)$, respectively.
These seven equations given in Eqs.~(\ref{eqn:gap})-(\ref{eqn:pO}) correspond to
our self-consistency equations at the mean-field level.
In the forthcoming Secs.~\ref{sec:bcs} and~\ref{sec:fflo}, we solve them to obtain 
the phase diagram of the system for the BCS- and FFLO-like superfluid phases, 
respectively.

\section{BCS-like uniform superfluid phases} 
\label{sec:bcs}
To analyze the BCS-like uniform superfluid phases, it is sufficient to solve four equations only, 
i.e. (\ref{eqn:gap}), (\ref{eqn:ntot}), (\ref{eqn:pI}) and (\ref{eqn:pO}), 
with $\mathbf{Q} = \mathbf{0}$, for a 
self-consistent set of  $|\Delta_0|, \mu, h_I$ and $h_O$ values. 
We also checked the stability of our mean-field solutions for the uniform superfluid 
phase using the curvature criterion, i.e. a nonuniform superfluid phase 
(e.g. a phase separation) is favored when $\partial^2 \Omega / \partial |\Delta_0|^2 < 0$,
since this criterion coincides with that of the compressibility one, i.e. the matrix 
$\kappa_{\sigma\sigma'} = -\partial^2 \Omega/(\partial \mu_\sigma \partial \mu_{\sigma'})$ 
must be positive-definite for the stability of the obtained solutions.
Before, we present the resultant phase diagrams, let us first analyze the 
excitation spectrum of the system.

\subsection{Excitation Spectrum} 
\label{sec:es}
In general, all four branches of the excitation spectrum may be different from each other, 
and hence both $E_{\mathbf{k 0}, \lambda}$ and their $|\Delta_0|, \mu, h_I$ and $h_O$ 
derivatives are not analytically tractable and need to be evaluated simultaneously with the 
self-consistency Eqs.~(\ref{eqn:gap}) and~(\ref{eqn:ntot}) for all $\mathbf{k}$-space points. 
This makes the current work numerically more demanding compared to the previous 
ones~\cite{gong, subasi, subasi2, wyi, carlos, zhang,liao,zhou,he}, for which the explicit 
forms of $E_{\mathbf{k 0}, \lambda}$ can be found. In order to understand the topology of the 
resultant superfluid phases, next we analyze the analytically-tractable limits.

\subsubsection{No SOC limit} 
In the absence of a SOC, setting $\mathbf{Q} = \mathbf{0}$ and $\alpha = 0$ in Eq.~(\ref{eqn:ham}), 
we obtain 
\begin{align}
E_{\mathbf{k 0},\lambda} = p_\lambda h_T + s_\lambda \sqrt{\xi_{\mathbf{k}}^2 + |\Delta_0|^2},
\end{align}
where $s_1 = s_2 = p_2 = p_3 = +1$ and $s_3 = s_4 = p_1 = p_4 =-1$ and $h_T = \sqrt{h_O^2 + h_I^2}$
is the strength of the total Zeeman field, showing that 
$E_{\mathbf{k 0},1} = - E_{\mathbf{k 0},3}$ can have zero-energy regions in $\mathbf{k}$ space.
The doubly-degenerate zeros are determined by $h_T^2 - \xi_{\mathbf{k}}^2 = |\Delta_0|^2$, 
and when $h_T > |\Delta_0|$ these conditions give two spheres of zeros for $\mu > \sqrt{h_T^2 - |\Delta_0|^2}$ 
and one sphere of zeros for $\mu < \sqrt{h_T^2 - |\Delta_0|^2}$. Therefore, the transition from 
gapped superfluid to gapless superfluid occurs at $h_T = |\Delta_0|$, and $\mu = 0$ 
determines the transition from gapless superfluid with two doubly-degenerate spheres 
of zeros to the one with one doubly-degenerate sphere of zeros, i.e. it gives the critical 
point for the disappearance of the doubly-degenerate inner sphere. 
These possibilities are schematically illustrated in Figs.~\ref{fig:topclass}(a) and~\ref{fig:topclass}(e), 
and the corresponding gapless superfluid phases in this case are known to be topologically 
trivial~\cite{gong, subasi, subasi2, wyi, carlos, zhang, liao,zhou,he, iskin}.

\begin{widetext} 
\begin{center}
(Figure~\ref{fig:nkbcs})
\end{center}
\begin{figure} [htb]
\centerline{\scalebox{0.36}{\includegraphics{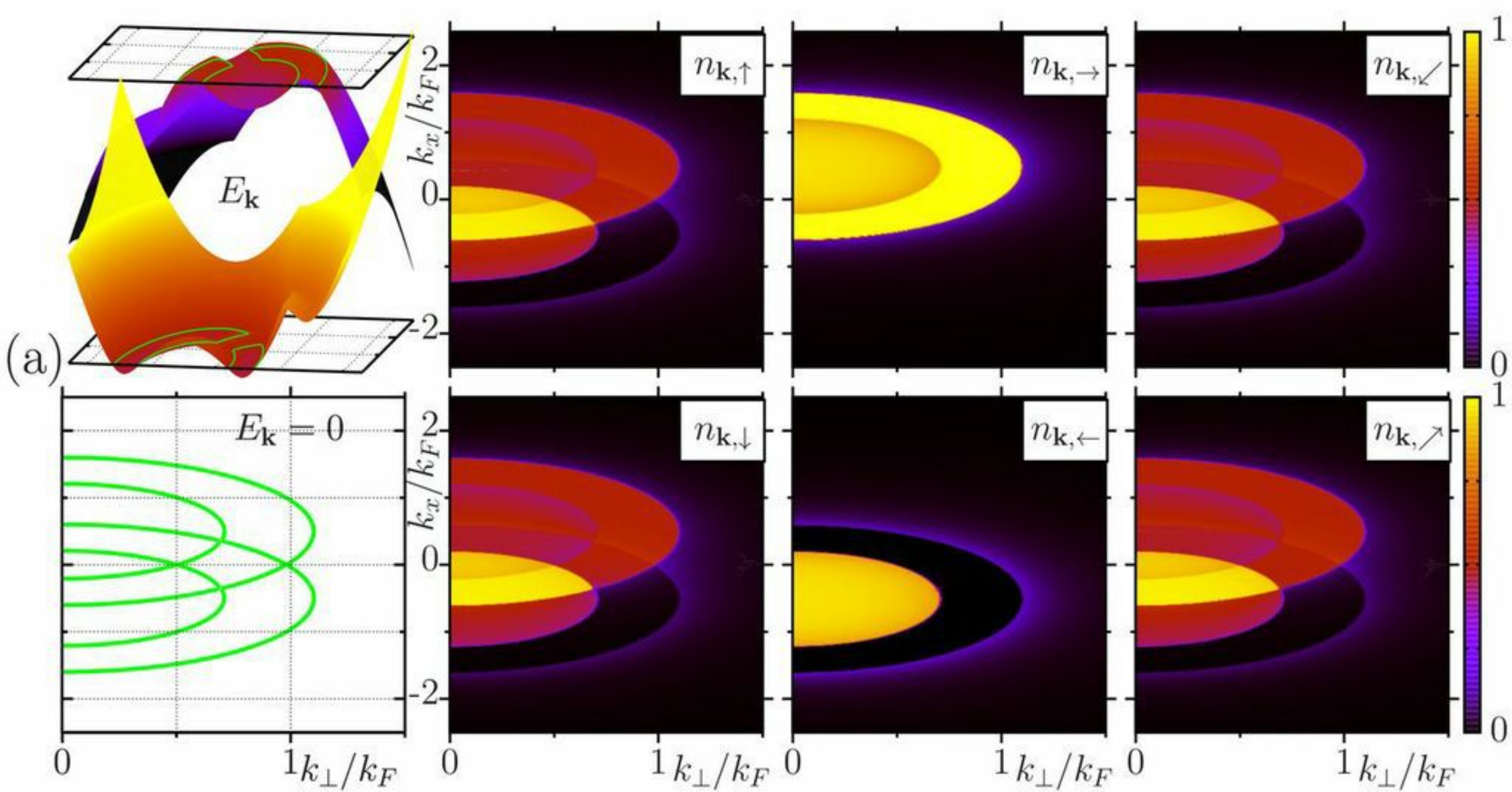}}}
\centerline{\scalebox{0.36}{\includegraphics{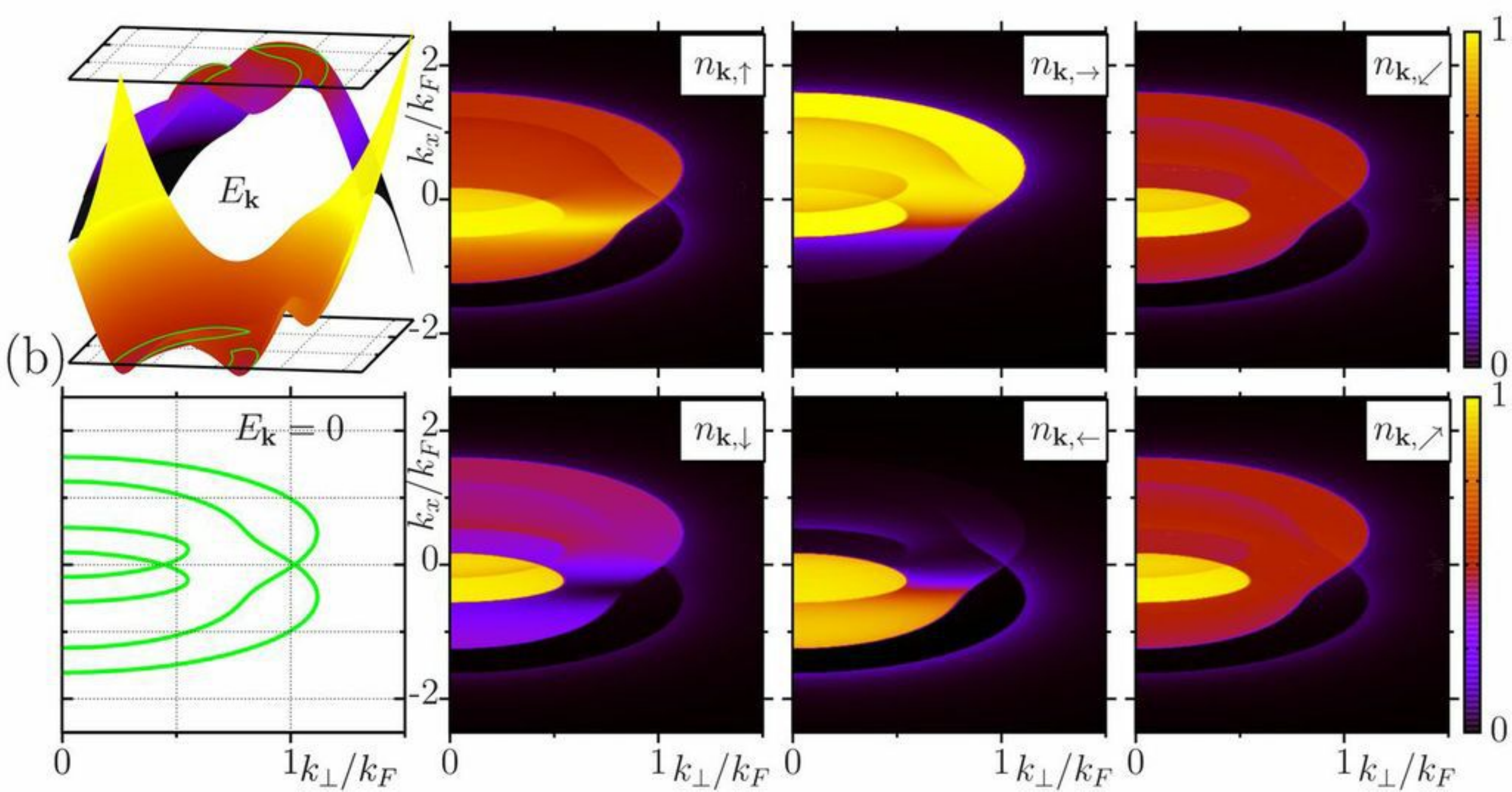}}}
\centerline{\scalebox{0.36}{\includegraphics{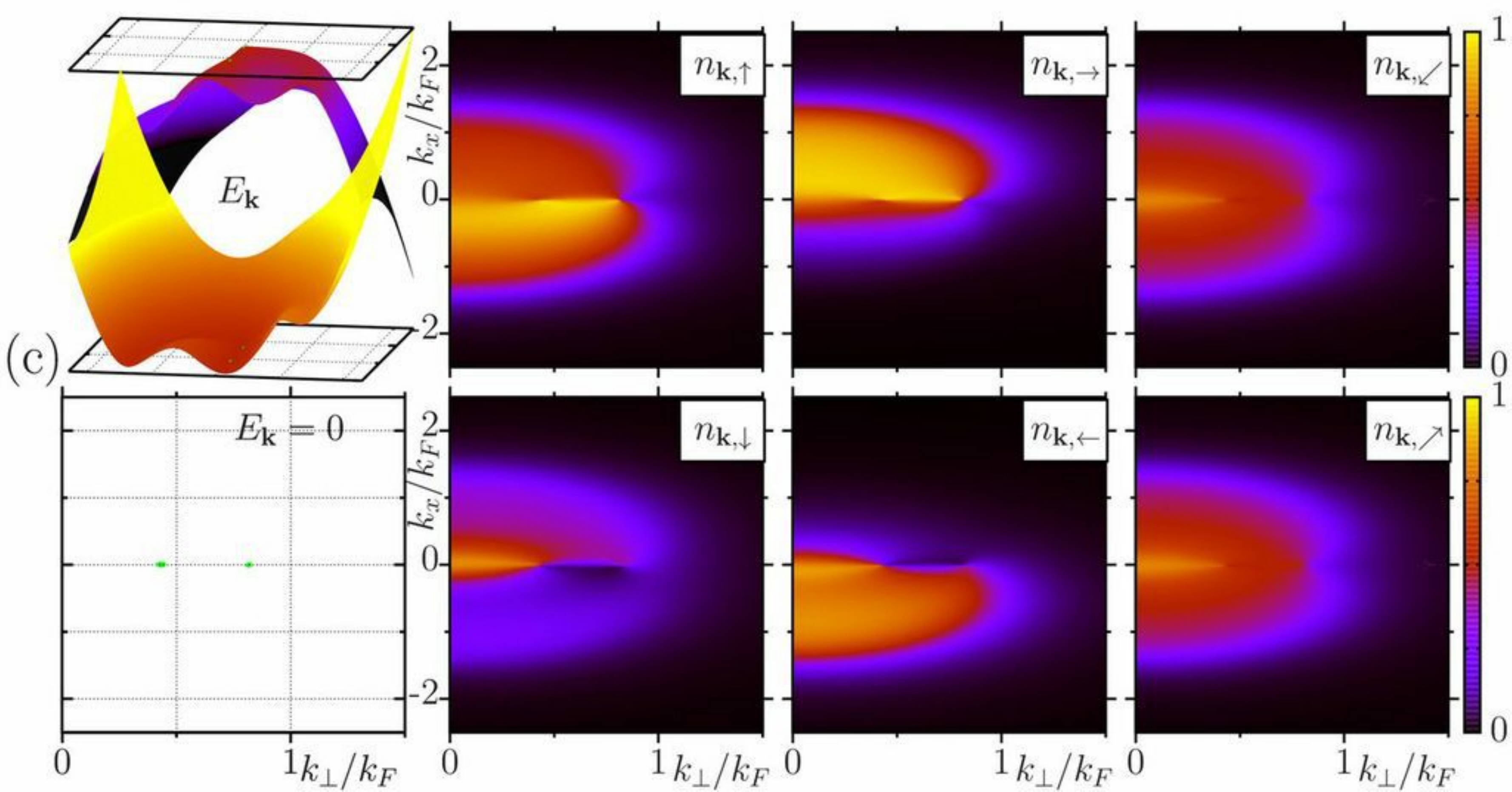}}}
\caption{
\label{fig:nkbcs} (color online)
The gapless branches of the energy spectrum $E_{\mathbf{k 0},1}$ and $E_{\mathbf{k 0},3}$ 
(in units of $\epsilon_F$) and the projections of the momentum distribution
$n_{\mathbf{k}, \delta}$, where $\delta = (\uparrow, \downarrow)$ for the out-of-plane $z$,
$\delta = (\rightarrow, \leftarrow)$ for the in-plane $y$ and 
$\delta = (\swarrow, \nearrow)$ for the $x$ direction,
are shown as a function of 
$k_x$ and $k_\perp = \sqrt{k_y^2+k_z^2}$ for the BCS-like uniform superfluid phases 
with (a) four spheres (b) four concave spheroids and (c) two doubly-degenerate 
circles of zeros. Here, we set $1/(k_F a_s) = 0$ and 
$\alpha = 0.5 k_F/M$, and choose $P_O = 0$ and $P_I = 0.5$ in (a), 
$P_O = 0.25$ and $P_I = 0.5$ in (b), and $P_O = 0.25$ and $P_I = 0.1$ in (c).
Note that $E_{\mathbf{k 0}, 1} = -E_{\mathbf{-k 0}, 3} = 0$ regions coincide with the 
locations of the sharp $n_{\mathbf{k},\delta}$ features.
}
\end{figure}
\end{widetext}

\subsubsection{No out-of-plane Zeeman field limit} 
In the absence of an out-of-plane Zeeman field, setting $\mathbf{Q} = \mathbf{0}$ and $h_O = 0$, 
we obtain 
\begin{align}
E_{\mathbf{k 0},\lambda} = p_\lambda h_I + s_\lambda \sqrt{(\xi_{\mathbf{k}} + p_\lambda \alpha k_x)^2 + |\Delta_0|^2},
\end{align}
showing that $E_{\mathbf{k 0},1} = -E_{-\mathbf{k 0},3}$ can have zeros. 
The zeros are determined by
$
\widetilde{k}_x^2 + k_\perp^2 = 2M(\widetilde{\mu} \pm \sqrt{h_I^2 - |\Delta_0|^2}),
$
where $\widetilde{k}_x = k_x - M\alpha$, $k_\perp = \sqrt{k_y^2 + k_z^2}$ and $\widetilde{\mu} = \mu - M\alpha^2/2$, 
and when $h_I > |\Delta_0|$ these conditions give four spheres of zeros for $\widetilde{\mu} > \sqrt{h_I^2 - |\Delta_0|^2}$ 
and two spheres of zeros for $\widetilde{\mu} < \sqrt{h_I^2 - |\Delta_0|^2}$. Therefore, the transition 
from gapped superfluid to gapless superfluid occurs at $h_I = |\Delta_0|$, and $\widetilde{\mu} = 0$ determines 
the transition from gapless superfluid with four spheres of zeros to the one with two spheres 
of zeros, i.e. it gives the critical point for the simultaneous disappearance of two inner (one each from 
$E_{\mathbf{k 0},1}$ and $E_{\mathbf{k 0},3}$) spheres. The superfluid phase with 
four spheres of zeros is illustrated in Fig.~\ref{fig:nkbcs}(a). In this figure, the main effect 
of ERD coupling $-\alpha k_x \sigma_y$ is best seen in the $y$ projection 
$n_{\mathbf{k}, \rightarrow}$ and $n_{\mathbf{k}, \leftarrow}$ of the momentum 
distribution as a $k_x$-dependent Zeeman field, since $h_y \ne 0$ breaks the 
time-reversal symmetry for this projection.
Note that since the SOC gauge field can be integrated out via a shift in the $k_x$ momentum 
and $\mu$, the corresponding gapless superfluid phases have the same topology 
and number as case (i), and therefore they are also topologically trivial. However, in comparison 
to case (i), since the SOC breaks the inversion symmetry in the $k_x$ direction, $\alpha \ne 0$ removes 
most of the degeneracy of zeros except for one or four circles, as schematically illustrated in 
Figs.~\ref{fig:topclass}(b) and~\ref{fig:topclass}(f). Note also that sufficiently large $\alpha$ 
splits these spheres of zeros completely away from each other, leaving no degeneracy.

\subsubsection{No in-plane Zeeman field limit} 
In the absence of an in-plane Zeeman field, setting $\mathbf{Q} = \mathbf{0}$ and $h_I = 0$, 
we obtain 
\begin{align}
E_{\mathbf{k 0},\lambda} = s_\lambda \sqrt{\xi_{\mathbf{k}}^2 + h_O^2 + \alpha^2 k_x^2 + |\Delta_0|^2 + 2s_\lambda p_\lambda A_{\mathbf{k}}},
\end{align}
where $A_{\mathbf{k}} = \sqrt{h_O^2(\xi_{\mathbf{k}}^2 + |\Delta_0|^2) + \alpha^2 k_x^2 \xi_{\mathbf{k}}^2}$,
showing that $E_{\mathbf{k 0},1} = - E_{\mathbf{k 0},3}$ can have zeros.
The doubly-degenerate zeros are determined by $k_x = 0$ and $h_O^2 - \xi_{\mathbf{k}}^2 = |\Delta_0|^2$, 
and when $h_O > |\Delta_0|$ these conditions give two circles of zeros for $\mu > \sqrt{h_O^2 - |\Delta_0|^2}$ 
and one circle of zeros for $\mu < \sqrt{h_O^2 - |\Delta_0|^2}$. Therefore, the transition from gapped 
superfluid to gapless superfluid occurs at $h_O = |\Delta_0|$, and $\mu = 0$ determines the transition from 
gapless superfluid with two doubly-degenerate circles of zeros to the one with one doubly-degenerate 
circle of zeros, i.e. it gives the critical point for the disappearance of the doubly-degenerate inner circle. 
These possibilities are shown in Figs.~\ref{fig:topclass}(d) and~\ref{fig:topclass}(h), and in sharp contrast 
to the cases (i) and (ii), the corresponding gapless superfluid phases in this case are known to be 
topologically nontrivial~\cite{subasi2, carlos}. The superfluid phase with 
two doubly-degenerate circles of zeros is also illustrated in Fig.~\ref{fig:nkbcs}(c),
where the main effect of ERD coupling is best seen in the $z$ and $y$ projections of 
the momentum distribution as a $k_x$-dependent Zeeman field. Note here that 
even though we choose $P_I \ne 0$ in this figure, as shown in Fig.~\ref{fig:pdbcs}(c), 
this particular data has the same $\mathbf{k}$-space topology as the $P_I = 0$ limit. 
Therefore, it is for this reason the time-reversal symmetry is broken by both $h_z\ne 0$ 
and $h_y \ne 0$, showing up in the $z$ and $y$ projections.

\subsubsection{The generic case} 
In the light of these limits, it is easier to understand the topology of zeros in the most 
general case when all three (Zeeman and SOC) gauge fields are nonzero. 
This generic case is schematically illustrated in Figs.~\ref{fig:topclass}(c) 
and~\ref{fig:topclass}(g), showing 
that $h_O \ne 0$ breaks the remaining degeneracy of two circles of 
zeros (only the ones at finite $k_x$) that are still present in Fig.~\ref{fig:topclass}(b). 
Therefore, the topology of gapless regions correspond to four concave 
spheroids on the BCS side and two concave spheroids on the BEC side, 
and the transition between the two occurs around $\mu \approx 0$, 
with the simultaneous disappearance of the two inner (one each from 
$E_{\mathbf{k 0},1}$ and $E_{\mathbf{k 0},3}$) concave spheroids. For instance, 
Fig.~\ref{fig:nkbcs}(b) shows the full $\mathbf{k}$ dependence of the gapless branches
of the excitation spectrum and projections of the momentum distribution 
for the superfluid phase with four concave spheroids of zeros.
Note again that the main effect of ERD coupling is best seen in the $z$ and $y$ 
projections of the momentum distribution as a $k_x$-dependent Zeeman field, 
since the time-reversal symmetry is broken by both $h_z \ne 0$ and $h_y \ne 0$ 
for these projections.
It is important to emphasize that the exotic superfluid phases discussed above, in particular 
the ones illustrated in Figs.~\ref{fig:topclass}(b),~\ref{fig:topclass}(c),~\ref{fig:topclass}(f) 
and~\ref{fig:topclass}(g), as well as the numerous possible 
ways of transitions in between are unique to this work. To explore the feasibility of observing 
these phases in cold-atom experiments, next we analyze the phase diagram of the system
at $T = 0$.

\subsection{Ground-state Phase Diagrams}
\begin{figure} [htb]
\centerline{\scalebox{0.63}{\includegraphics{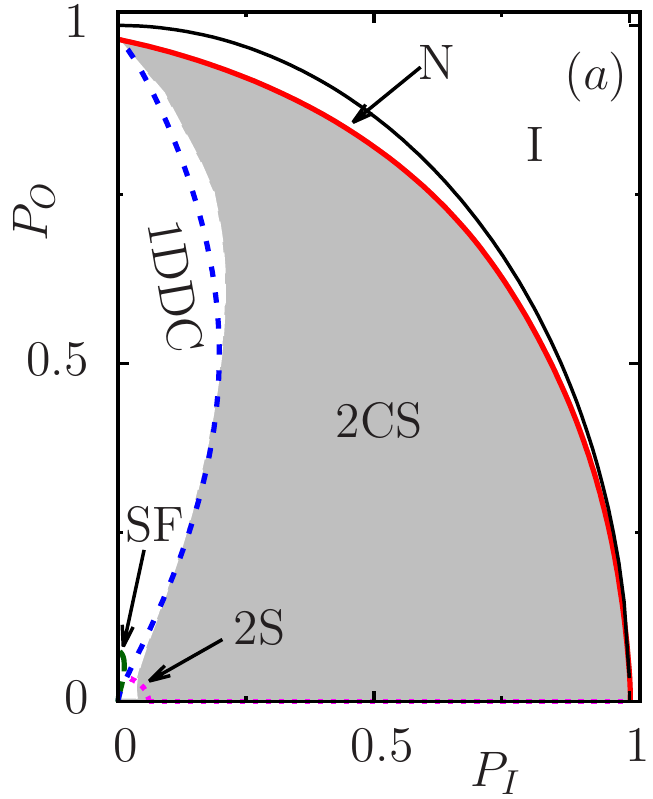}
\includegraphics{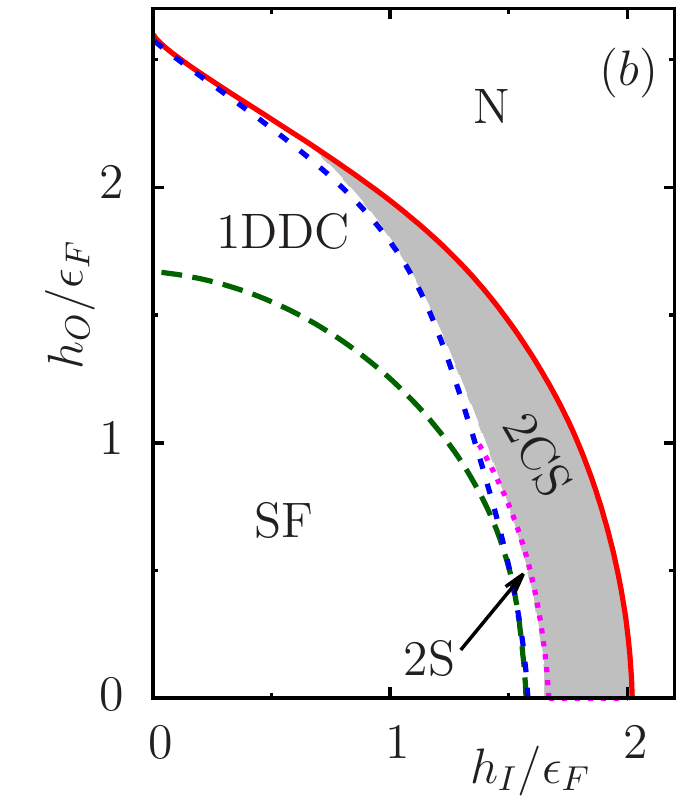}}}
\centerline{\scalebox{0.63}{\includegraphics{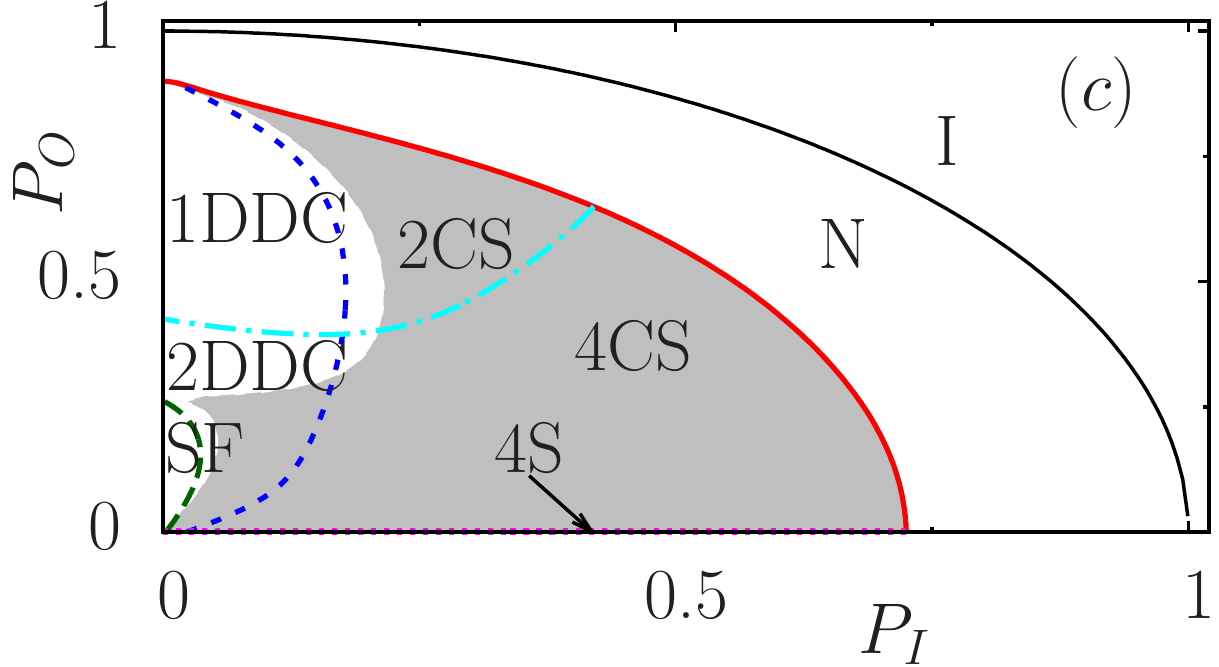}}}
\caption{
\label{fig:pdbcs} (color online)
The ground-state phase diagram is shown for the BCS-like uniform superfluid phases
as a function of out-of- and in-plane (a,c) number polarizations and (b) Zeeman fields, 
where $1/(k_F a_s) = 1$ in (a, b) and 0 in (c). Here, $\alpha = 0.5 k_F/M$ in all figures, 
and the labels correspond to an Inaccessible region (I), normal (N) phase, gapped superfluid 
(SF), and gapless superfluid with zeros of one or two doubly-degenerate circles (1DDC, 2DDC),
two or four spheres (2S, 4S), or two or four concave spheroids (2CS, 4CS). 
The curvature criterion is $\partial^2 \Omega / \partial |\Delta_0|^2 < 0$ in the shaded region,
indicating an instability toward a nonuniform superfluid phase.
}
\end{figure}

The generic phase diagrams are shown in Fig.~\ref{fig:pdbcs}.
We note that, since the SOC can be integrated out of the self-consistency 
Eqs.~(\ref{eqn:gap}) and~(\ref{eqn:ntot}) on the $x$ axis when $h_O = 0$, 
the critical $h_I$ or $P_I$ for the transitions from a gapped superfluid to a gapless 
one with two or four spheres of zeros and from the latter to the normal phase 
are exactly the same as those obtained for a population-imbalanced Fermi gas 
without the SOC, i.e. for the transition from a gapped superfluid to a gapless one 
with one or two doubly-degenerate spheres of zeros and from the latter to 
the normal phase. Here, we locate the normal phase boundaries by finding 
$|\Delta_0| \lesssim 10^{-3} \epsilon_F$ where $\epsilon_F = k_F^2/(2M)$ is 
the Fermi energy, and the inaccessible regions are determined 
by $P_I^2 + P_O^2 \ge 1$. In addition, on the $y$ axis when $h_I = 0$, our results 
recover the recent works~\cite{subasi2, carlos}.

Away from the $x$ and $y$ axes, we find that the gapless superfluid phases with two
or four concave spheroids of zeros occupy quite large regions in the phase diagrams 
as shown in Fig.~\ref{fig:pdbcs}. Depending on $\alpha$ and $a_s$,
note that the transition from a gapless superfluid phase with two or four spheres 
of zeros to the one with two or four concave spheroids of zeros may require a finite 
threshold for $h_O$. For instance, when the SOC is weak so that the spheres of 
zeros contain doubly-degenerate circles as shown in Figs.~\ref{fig:topclass}(b) 
and~\ref{fig:topclass}(f), an arbitrarily small $h_O$ splits these degeneracies, 
immediately giving rise to the transition. This is clearly seen in Fig.~\ref{fig:pdbcs}(c). 
On the contrary, when the SOC is sufficiently strong so that it splits the spheres 
of zeros completely away from each other leaving no degeneracy, the transition 
occurs at a finite $h_O$, as shown in Figs.~\ref{fig:pdbcs}(a) and~\ref{fig:pdbcs}(b). 

Next we note an important difference between the grand-canonical (Zeeman fields) 
and canonical (number polarizations) phase diagrams on the BCS side of the 
BCS-BEC evolution, and consider the $h_O = 0$ limit for the sake of its simplicity. 
In a grand-canonical phase diagram, where $h_I$ is increased from 0, it can be shown that 
while $|\Delta_0|$ is unaffected by $h_I \ne 0$ until $h_I = |\Delta_{00}|$, where $|\Delta_{00}|$ 
is the value of the field-free order parameter when $h_I = 0$, it immediately vanishes for 
$h_I > |\Delta_{00}|$. Therefore, increasing $h_I$ leads to a first-order quantum phase 
transition between a gapped superfluid and normal phase, without any intermediate 
gapless superfluid phase in between. On the other hand, in the case of a canonical 
phase diagram, where $P_I$ is increased from 0, there exists a $|\Delta_0| \ne 0$ solution 
with $h_c < h_I < |\Delta_{00}|$, where $h_c$ is determined by the maximum 
possible number polarization $P_I = P_c$. Therefore, in sharp contrast to the grand-canonical case, 
it is possible to obtain a gapless superfluid phase with increasing $P_I$. 
For this reason, the canonical phase diagram suits better for presenting 
the possible gapless superfluid phases at unitarity as shown in Fig.~\ref{fig:pdbcs}(c). 
However, on the BEC side of the BCS-BEC evolution, such a complication does 
not exist, and it is possible to map the grand-canonical and canonical phase diagrams 
onto each other as shown in Figs.~\ref{fig:pdbcs}(a) and~\ref{fig:pdbcs}(b). 

We emphasize that even though all of these gapless superfluid phases have order parameters
with the same $s$-wave symmetry, they can be classified with respect to the topology 
of their zeros. It is well-known in other contexts that such changes in the topology 
leave strong evidences on the thermodynamic quantities such as the atomic compressibility, 
spin susceptibility and momentum distribution, signaling a topological quantum 
phase transition. For instance, in nodal $p$-wave superconductors and superfluids 
for which the order parameters are $\mathbf{k}$ dependent, while the excitation 
spectrum $E_{\mathbf{k}} = \sqrt{\xi_\mathbf{k}^2 + |\Delta_\mathbf{k}|^2}$ has gapless 
points of zeros in $\mathbf{k}$ space (which are determined by the nodes of
$|\Delta_\mathbf{k}| = 0$) when $\mu > 0$ on the BCS side, it is fully gapped when 
$\mu < 0$ on the BEC side, showing a Lifshitz-type phase transition in the 
BCS-BEC evolution which occurs precisely at $\mu = 0$~\cite{volovik, read}. 
The presented phase diagrams clearly demonstrate here that it is possible to simulate 
a much richer variety of gapless superfluid phases and numerous possible 
ways of phase transitions between them with active experiments. Next, we 
investigate whether these phases survive if we allow a finite center-of-mass 
momentum $\mathbf{Q}$.

\section{FFLO-like nonuniform superfluid phases} 
\label{sec:fflo}
In the previous section, having shown that the presence of an in-plane Zeeman 
field has a dramatic effect on the stability of BCS-like uniform superfluid phases 
with $\mathbf{Q} = \mathbf{0}$, i.e. nonuniform superfluid phases occupy 
a very large region in the phase diagrams when $P_I \ne 0$, here we discuss 
the possibility of FFLO-like nonuniform (spatially-modulated) superfluid phases with 
$\mathbf{Q} \ne \mathbf{0}$ as a possible ground state of the system such that
$\Delta_{\mathbf{Q}} = |\Delta_0| e^{i\mathbf{Q} \cdot \mathbf{R}}$ (see Sec.~\ref{sec:mft}). 
For this purpose, one needs to solve all seven equations, i.e. Eqs.~(\ref{eqn:gap})-(\ref{eqn:pO}), 
in general for a self-consistent set of  $|\Delta_0|, Q_x, Q_y, Q_z, \mu, h_I$ and $h_O$ 
values, which is a highly nontrivial task. Our numerical solutions suggest that 
$Q_x \gg \{Q_y, Q_z\} \to 0$ for a greater portion of the parameter space, and therefore, 
we set $Q_y = Q_z = 0$, and solve only for $Q_x$.

\begin{figure} [htb]
\centerline{\scalebox{0.63}{\includegraphics{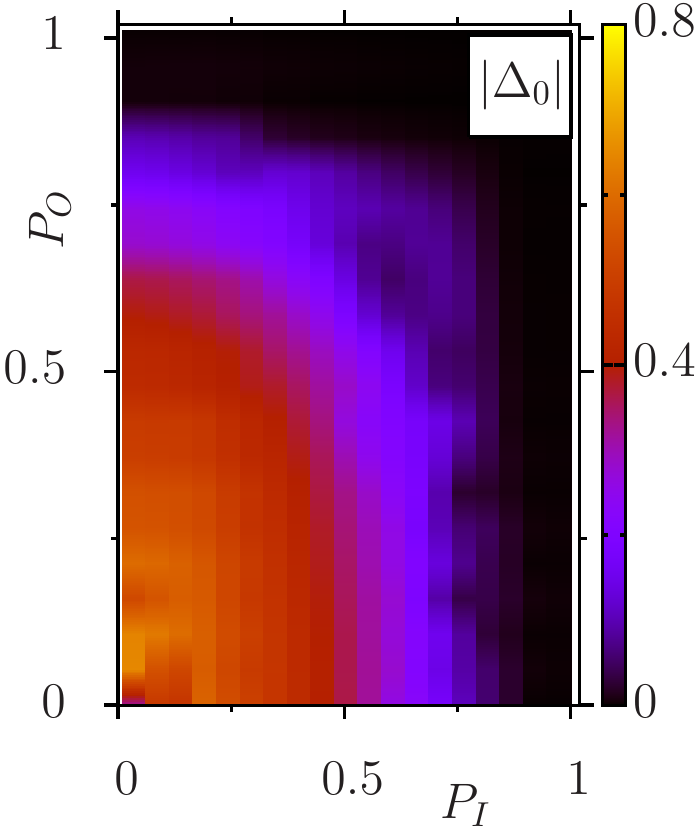}
\includegraphics{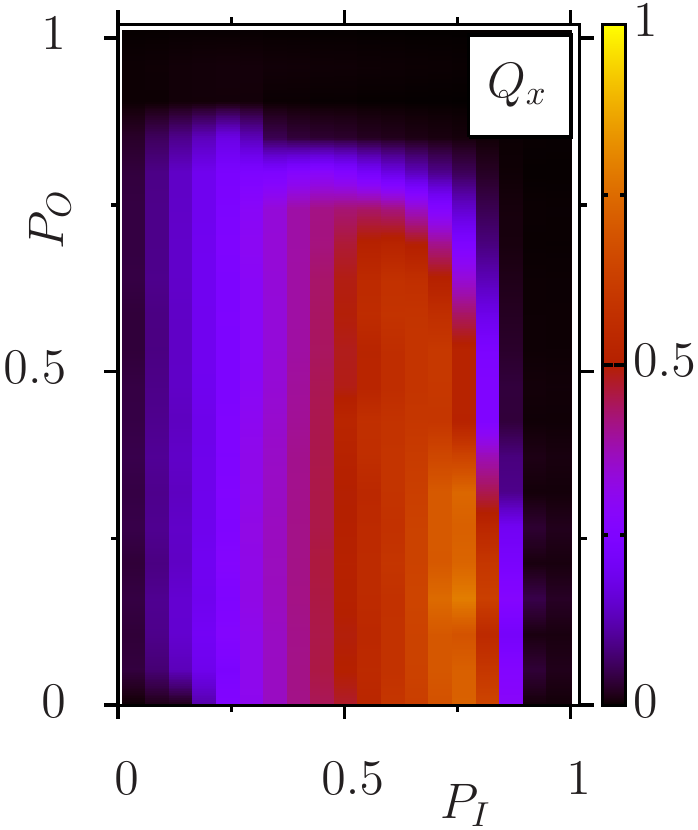}}}
\centerline{\scalebox{0.63}{\includegraphics{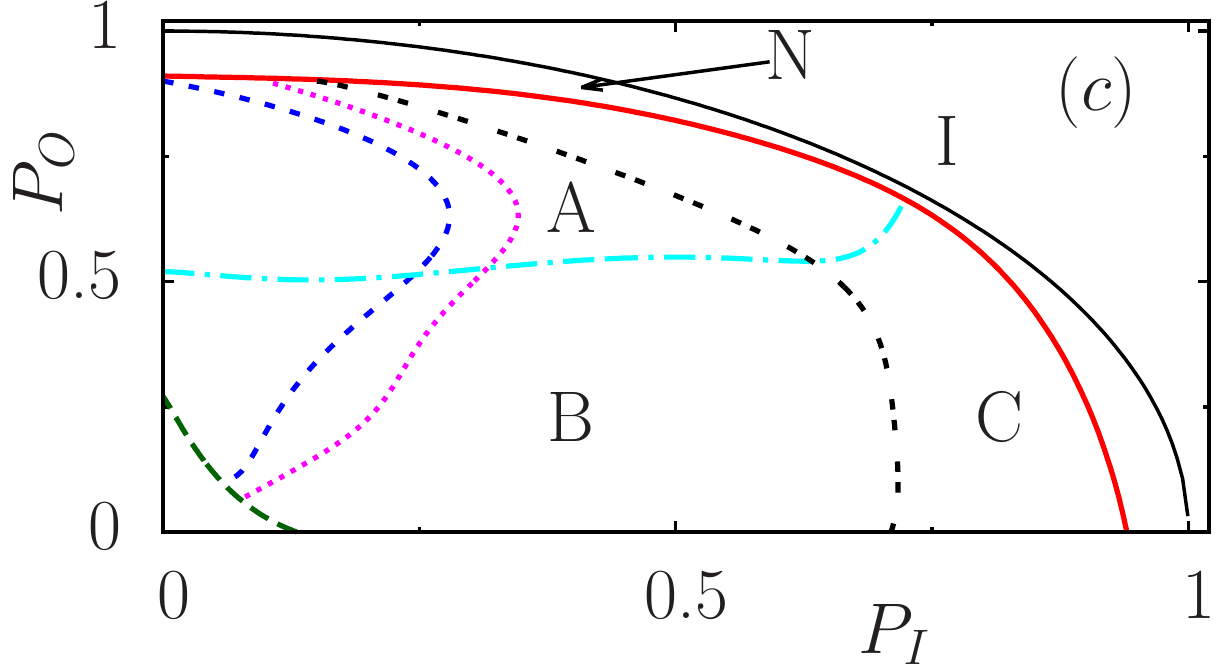}}}
\caption{
\label{fig:pdfflo} (color online)
The color maps of the magnitude $|\Delta_0|$ (in units of $\epsilon_F$) 
of the order parameter $\Delta_\mathbf{Q} = |\Delta_0| e^{i \mathbf{Q} \cdot \mathbf{R}}$
and $x$ component $Q_x$ (in units of $k_F$) of the center-of-mass momentum,
and the ground-state phase diagram are shown for the FFLO-like nonuniform 
superfluid phases,  where we set $1/(k_F a_s) = 0$ and $\alpha = 0.5 k_F/M$, 
i.e. same as in Fig.~\ref{fig:pdbcs}(c). The FFLO-like phases are further
classified with respect to the topology of the momentum-space 
regions with zero excitation energy in (c). The $A$, $B$ and $C$ regions are 
further illustrated in Fig.~\ref{fig:nkfflo}.
}
\end{figure}

In Fig.~\ref{fig:pdfflo}, we show the color maps of the magnitude of the order
parameter $|\Delta_0|$ and $x$ component $Q_x$ of the center-of-mass momentum, 
together with the ground-state phase diagram when $1/(k_F a_s) = 0$ and 
$\alpha = 0.5 k_F/M$. These parameters correspond to that of the figure shown 
in Fig.~\ref{fig:pdbcs}(c). First of all, we find that $Q_x = 0$ on the $P_O$ axis when 
$P_I = 0$, and therefore, conclude that FFLO-like nonuniform superfluid phases 
are not favored in the absence of an in-plane Zeeman field, which is consistent with 
the stability analysis given in Fig.~\ref{fig:pdbcs}(c). On the other hand, we find 
that $Q_x$ grows almost linearly on the $P_I$ axis, i.e. $Q_x/k_F \simeq P_I$, 
and the FFLO-like phases are strongly favored in the absence of an 
out-of-plane Zeeman field, which is again consistent with the stability 
analysis given in Fig.~\ref{fig:pdbcs}(c). Therefore, there is a 
clear competition between out-of- and in-plane Zeeman fields when both fields 
are present. It turns out that the effects of in-plane field rapidly dominates 
over the out-of-plane one, as a result of which FFLO-like phases 
occupy almost the entire phase diagram. In addition, a direct comparison 
between Figs.~\ref{fig:pdbcs}(c) and~\ref{fig:pdfflo}(c) shows that FFLO-like 
phases are more favored than the BCS-like ones against the normal phase.

\begin{widetext} 
\begin{center}
(Figure~\ref{fig:nkfflo})
\end{center}
\begin{figure} [htb]
\centerline{\scalebox{0.36}{\includegraphics{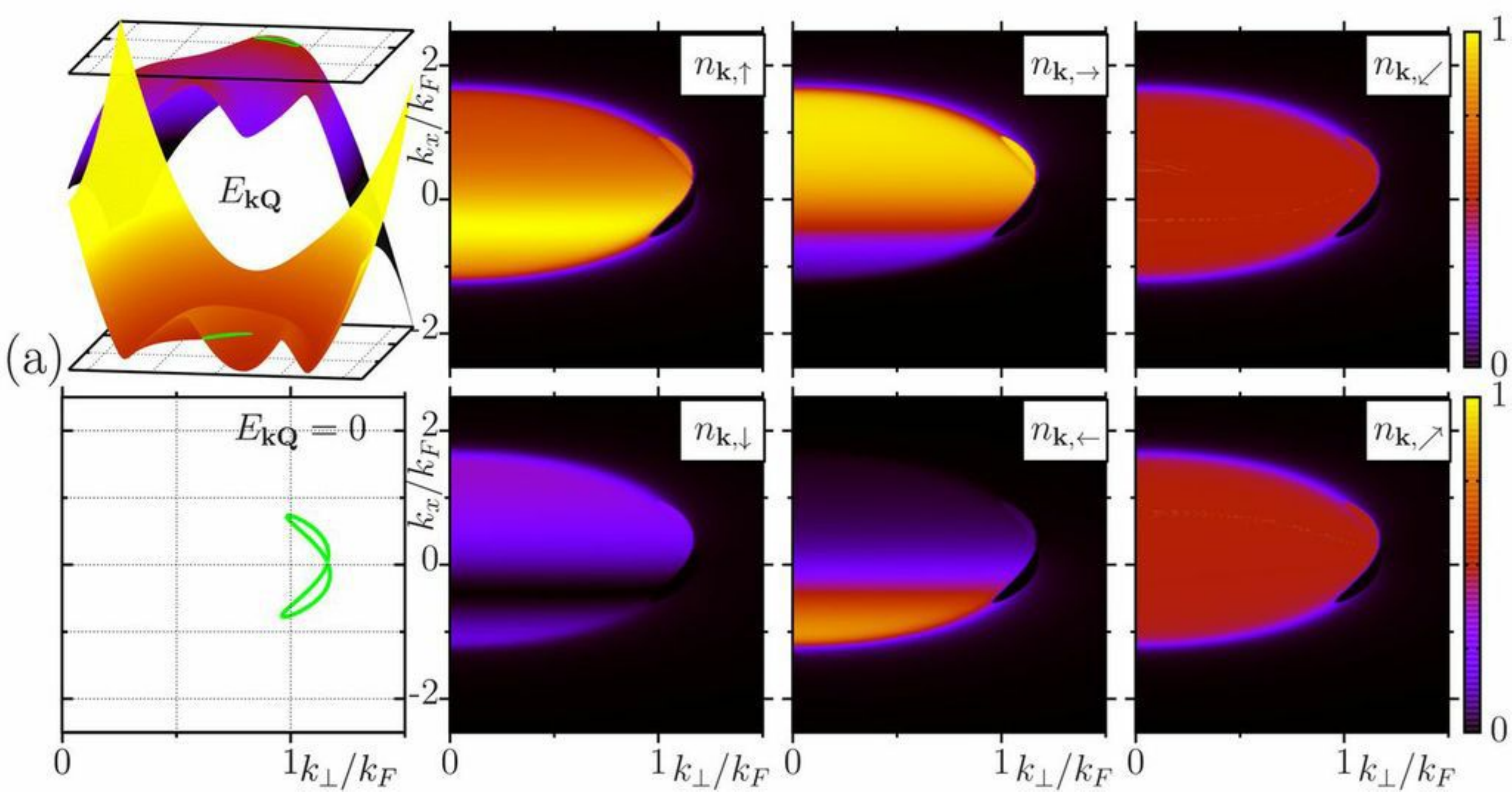}}}
\centerline{\scalebox{0.36}{\includegraphics{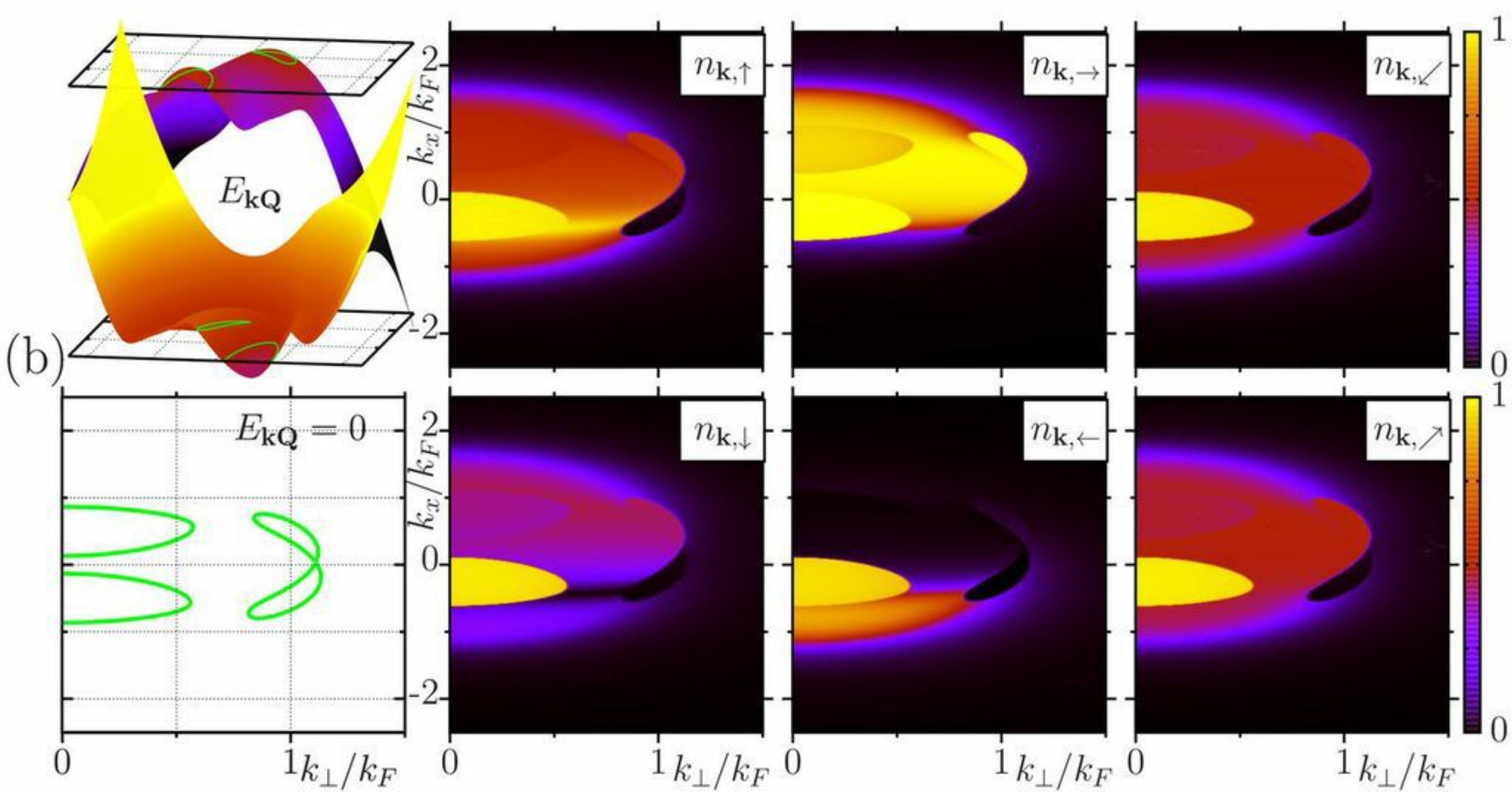}}}
\centerline{\scalebox{0.36}{\includegraphics{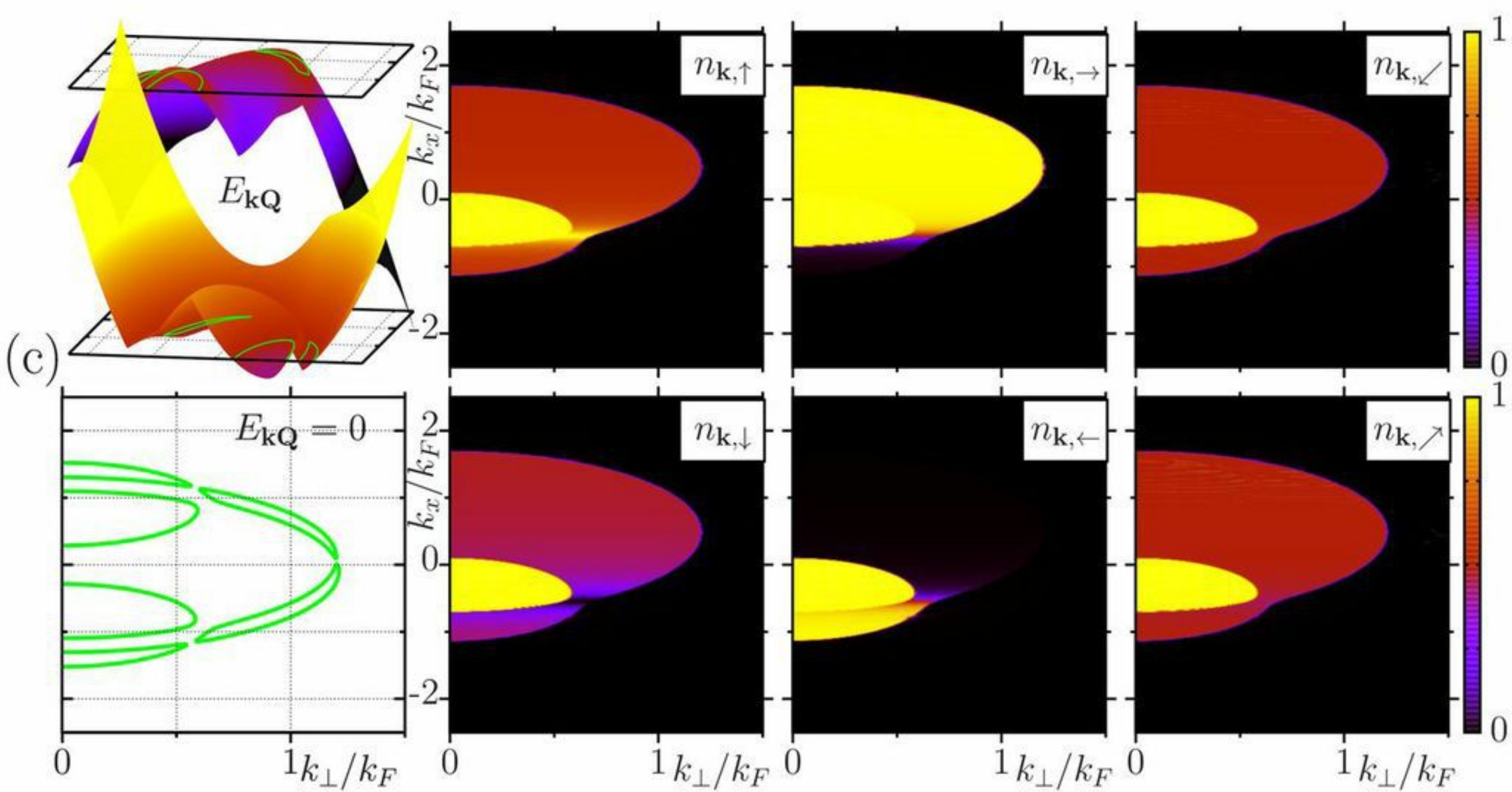}}}
\caption{
\label{fig:nkfflo} (color online)
The gapless branches of the energy spectrum $E_{\mathbf{k Q},1}$ and $E_{\mathbf{k Q},3}$ 
(in units of $\epsilon_F$) and the projections of the momentum distribution 
$n_{\mathbf{k}, \delta}$, where $\delta = (\uparrow, \downarrow)$ for the out-of-plane $z$,
$\delta = (\rightarrow, \leftarrow)$ for the in-plane $y$ and 
$\delta = (\swarrow, \nearrow)$ for the $x$ direction, 
are shown as a function of $k_x$ and $k_\perp = \sqrt{k_y^2+k_z^2}$ 
for the FFLO-like nonuniform superfluid phases.
Here, we set $1/(k_F a_s) = 0$ and 
$\alpha = 0.5 k_F/M$, and choose $P_O = 0.6$ and $P_I = 0.5$ in (a), 
$P_O = 0.2$ and $P_I = 0.5$ in (b), and $P_O = 0.1$ and $P_I = 0.75$ in (c),
and they represent, respectively, the typical topology of $A$, $B$ and $C$ 
regions of Fig.~\ref{fig:pdfflo}. Note that $E_{\mathbf{k Q}, 1} = -E_{\mathbf{-k Q}, 3} = 0$ 
regions and the locations of the sharp $n_{\mathbf{k},\delta}$ features are 
shifted by $\mathbf{Q}/2$.
}
\end{figure}
\end{widetext}

We note that while allowing the superfluid order parameter to have a single 
nonzero center-of-mass momentum $\mathbf{Q}$ 
favors the nonuniform phase over the uniform one where $\mathbf{Q} = \mathbf{0}$,
this may not be the case if we allow multiple $\mathbf{Q}$ momenta, 
e.g. $\Delta_{\mathbf{Q}} = |\Delta_0| \cos(\mathbf{Q} \cdot \mathbf{R})$.
This is because FFLO-like nonuniform superfluid phases 
in spin-orbit coupled Fermi gases are stabilized mainly by the asymmetry of the 
Fermi surfaces in $\mathbf{k}$ space, and this mechanism is in contrast with 
that of the condensed-matter ones where they are stabilized by the symmetric 
Zeeman mismatch in $\mathbf{k}$ space. Therefore, only in the latter case, we 
expect the solutions involving the superpositions of $\mathbf{Q}$ and $\mathbf{-Q}$ 
momenta to be favored even more than the $\mathbf{+Q}$ or $\mathbf{-Q}$ 
solutions themselves. However, such a comparison is beyond the scope of 
this work.

In addition, similar to our analysis for the BCS-like uniform superfluid phases 
given above in Sec.~\ref{sec:bcs}, we further classify the FFLO-like nonuniform 
superfluid phases with respect to the topology of the momentum-space 
regions with zero excitation energy. Although the zeros of some of the phases shown 
in Fig.~\ref{fig:pdfflo}(c) are very similar to the BCS-like ones shown in 
Figs.~\ref{fig:topclass} and~\ref{fig:pdbcs}(c), there are also a variety of new 
ones arising due solely to $\mathbf{Q} \ne 0$.
For instance, in Fig.~\ref{fig:nkfflo}, we illustrate the gapless branches of 
the excitation spectrum as well as the 
projections of the momentum distribution $n_{\mathbf{k},\delta}$ for three 
of the topologically distinct phases labeled in Fig.~\ref{fig:pdfflo}(c). 
Unlike the case of BCS-like superfluid phases shown in Fig.~\ref{fig:nkbcs}, note that
the particle-hole symmetry $E_{\mathbf{k Q}, 1} = -E_{\mathbf{-k Q}, 3} = 0$
does not necessarily require the zero-energy regions of the FFLO-like 
superfluids to have the inversion symmetry around $k_x = 0$. 
In addition, we see that the main effect of ERD coupling is best seen in the 
$z$ and $y$ projections of the momentum distribution as a $k_x$-dependent 
Zeeman field, since the time-reversal symmetry is broken by both $h_z \ne 0$ 
and $h_y \ne 0$ for these projections.
For FFLO-like superfluid phases, this figure clearly shows that sharp features 
of $n_{\mathbf{k},\delta}$ and zero-excitation-energy regions 
$E_{\mathbf{k Q}, 1} = -E_{\mathbf{-k Q}, 3} = 0$ do not match in $\mathbf{k}$-space, 
and they are shifted exactly by $\mathbf{Q}/2$, i.e. $Q_x/2$ in our paper. 
These shifts are important since it is in sharp contrast to the BCS-like 
superfluid phases shown in Fig.~\ref{fig:nkbcs}, 
where sharp features of $n_{\mathbf{k},\delta}$ and 
zero-excitation-energy regions exactly coincide in $\mathbf{k}$ space.
Therefore, the competing BCS-like gapless uniform and FFLO-like nonuniform 
superfluid ground states can easily be distinguished by directly looking at their 
momentum distributions, which has probably been the most commonly used 
probing technique in cold atom systems since the early days.

\section{Conclusions} 
\label{sec:conclusions}
In summary, we analyzed the effects of both in- and out-of-plane Zeeman fields 
on the BCS-BEC evolution of a Fermi gas with an ERD spin-orbit coupling field 
at zero temperature. Depending on the combination of these gauge fields, we found 
novel gapless superfluid phases that can be distinguished with respect to the 
topology of their zeros. For instance, for the BCS-like uniform superfluid phases 
with zero center-of-mass momentum (where $\Delta_{\mathbf{Q}} = \Delta_0$), 
the zeros may correspond to one or two 
doubly-degenerate spheres, two or four spheres, two or four concave 
spheroids, or one or two doubly-degenerate circles. We noted that these 
phases are distinct because, even though their order parameters all have the 
same $s$-wave symmetry, changes in the topology leave strong evidences 
on some of the thermodynamic quantities, signaling a topological 
quantum phase transition. 

We also studied the possibility of FFLO-like nonuniform (spatially-modulated) 
superfluid phases with finite center-of-mass momentum and obtained an 
even richer phase diagram.
In particular, we found that while FFLO-like phases are not favored against 
the BCS-like ones in the absence of an in-plane Zeeman field, 
even the simplest FFLO-like phases (where the superfluid order parameters 
involve only a single center-of-mass momentum $\mathbf{Q}$ in such a way that 
$\Delta_{\mathbf{Q}} = |\Delta_0| e^{i \mathbf{Q} \cdot \mathbf{R}}$)
are strongly favored in the absence of an out-of-plane Zeeman field. 
Therefore, there is a clear competition between 
out-of- and in-plane Zeeman fields when both fields are present, and it turns 
out that the effects of in-plane field rapidly dominates over the out-of-plane one, 
as a result of which even the simplest FFLO-like phases occupy almost the entire 
phase diagram. We also noted that allowing multiple center-of-mass momenta, 
e.g. $\Delta_{\mathbf{Q}} = |\Delta_0| \cos(\mathbf{Q} \cdot \mathbf{R})$, 
is not expected to favor and stabilize the FFLO-like phases much more than 
the single-$\mathbf{Q}$ case discussed in this paper. 
In addition, we further classified the FFLO-like phases with 
respect to the topology of the momentum-space regions with zero excitation 
energy. 

To conclude, our results suggest that cold atom systems with ERD coupling and 
in-plane Zeeman field are one of the best candidates for studies on exotic 
BCS-like gapless uniform and FFLO-like nonuniform superfluid phases, 
with a greater premise on a route toward studying them under highly controllable 
atomic settings. As a final remark, it is worth emphasizing that these phases 
can easily be distinguished by directly measuring their momentum distributions, 
which is routinely done in every cold atom system via simply looking at their
time-of-flight images.

\section{Acknowledgments}
This work is supported by the Marie Curie IRG (FP7-PEOPLE-IRG-2010-268239), 
T\"{U}B$\dot{\mathrm{I}}$TAK Career (3501-110T839), T\"{U}BA-GEB$\dot{\mathrm{I}}$P,
and the National Center for High Performance Computing of Turkey (UHeM-1001232011).

\end{document}